\documentstyle[12pt,aasms4]{article}

\def\logz{\lbrack\hbox{Fe/H}\rbrack}
\lefthead{Dolphin et al.}
\righthead{IC 1613 Variables}
\begin{document}

\title{Deep HST Imaging of IC 1613. I. Variable Stars and
Distance\linespread{1.0}\footnote{
Based on observations with the NASA/ESA \textit{Hubble Space Telescope},
obtained at the Space Telescope Science Institute, which is operated by
the Association of Universities for Research in Astronomy, Inc., under
NASA contract NAS 5-26555. These observations are associated with proposal
ID 7496.}}

\author{Andrew E. Dolphin, A. Saha}
\affil{National Optical Astronomy Observatories, P.O. Box 26372, Tucson, 
AZ 85726\\
Electronic mail: dolphin@noao.edu, saha@noao.edu}

\author{Evan D. Skillman}
\affil{Astronomy Department, University of Minnesota,
    Minneapolis, MN 55455\\
Electronic mail: skillman@astro.umn.edu}

\author{Eline Tolstoy}
\affil{UK Gemini Support Group, Astrophysics, NAPL,
 University of Oxford, Keble Road,
 Oxford OX1 3RH, UK \\
Electronic mail: etolstoy@astro.ox.ac.uk}

\author{A.A. Cole}
\affil{Physics \& Astronomy Dept., 538 LGRT,
University of Massachusetts, Amherst, MA 01003\\
Electronic mail: cole@condor.astro.umass.edu}

\author{R.C. Dohm-Palmer}
\affil{University of Michigan, Dept. of Astronomy, 821 Dennison Building,
Ann Arbor, MI 48109-1090\\
Electronic mail: rdpalmer@astro.lsa.umich.edu}

\author{J.S. Gallagher}
\affil{University of Wisconsin, Dept. of Astronomy, 475 N. Charter St,
Madison, WI 53706\\
Electronic mail: jsg@astro.wisc.edu}

\author{Mario Mateo}
\affil{University of Michigan, Dept. of Astronomy, 821 Dennison Building,
Ann Arbor, MI 48109-1090\\
Electronic mail: mateo@astro.lsa.umich.edu}

\author{J.G. Hoessel}
\affil{University of Wisconsin, Dept. of Astronomy, 475 N. Charter St,
Madison, WI 53706\\
Electronic mail: hoessel@astro.wisc.edu}

\begin{abstract}
We present WFPC2 $VI$ photometry of a field in the halo of IC 1613, finding
13 RR Lyraes and 11 Cepheids. Our photometry of the red giant branch tip
and red clump is used to derive distances to IC 1613, which are consistent
with each other and with distances based on the variable stars. We compare
these values with similarly-measured distances for the Magellanic Clouds,
and are able to measure metallicity dependencies of the RR Lyrae and Cepheid
distances by requiring consistent relative distance measurements from the
four techniques. For metallicities of $\logz = -1.3$
(RR Lyraes) and $-1.0$ (Cepheids), we find a relatively steep slope of
$0.34 \pm 0.20$ magnitudes per dex for the RR Lyraes and a shallow slope of
$-0.07 \pm 0.16$ mag/dex for the Cepheids, both values within the range
of theoretical and empirical results in the literature. We find that a
dependence of the red clump absolute magnitude on age, in addition to
metallicity, is required to produce self-consistent relative distances
between IC 1613 and the Magellanic Clouds. Adopting such a red clump
calibration and self-consistent calibrations for the other three distance
indicators, we find that the distances to all three objects are in
excellent agreement. Our best distance modulus to IC 1613 is
$\mu_0 = 24.31 \pm 0.06$,
corresponding to a distance of $730 \pm 20$  kpc. This distance produces
an RR Lyrae absolute magnitude of $M_V = 0.61 \pm 0.08$.
\end{abstract}


\section{Introduction}

Studies of the stellar populations in nearby galaxies provide a powerful 
tool for determining the key physical parameters of galaxy evolution 
such as the age (star formation history), the chemical composition and 
enrichment history, the stellar initial mass function, environmental 
effects, and the dynamical history of the system.  Using the HST,
it is possible to photometer individual stars down to very faint
magnitudes, and to interpret the observable parameters
such as the morphology of the color-magnitude diagram (CMD).
This approach is a logical stepping stone to understanding
galaxy evolution and provides a physical basis for understanding
observations of high redshift galaxies and their implications for
cosmological models.  Detailed analysis of the intermediate and old 
stellar populations of Local Group galaxies should reveal histories
in accordances with those implied by studies of galaxies at
higher redshift (Tolstoy 1999).  With the appropriate data, which currently
are obtained primarily by using HST, this hypothesis can be directly
tested.

Here we are presenting new HST observations of a field in the halo of
IC 1613, a dwarf 
irregular galaxy in the Local Group.  In this paper we concentrate
on the variable stars of IC 1613 and, in particular, their relevance
to its distance determination.  In a future paper, we will reconstruct
the star formation history of IC 1613 by analysis of the stellar
populations.

The known properties of IC 1613 have recently been summarized by van 
den Bergh (2000).  We will give only a short summary of relevant 
properties.  Because of its proximity (distance $\sim 720$ kpc), its
high galactic latitude ($-$60.6$^o$) and thus little galactic extinction,
and its inclination (38$^o$, Lake \& Skillman 1989), IC 1613 provides an
excellent opportunity to observe stellar populations in a relatively
low-metallicity environment.  Interestingly, there are few studies of
the metallicity in IC 1613. Mateo (1998) gives a value of the
mean iron abundance for the old and intermediate age stellar populations
of $\logz= -1.3 \pm 0.2$ (from Lee et al. 1993 using the RGB color at
$M_I = -3.5$; see also Cole et al. 1999) and an oxygen
abundance of the interstellar medium of 12 + log(O/H) $= 7.8 \pm 0.2$
(Talent 1980).  Due to the high Galactic latitude, the reddening
to IC 1613 is very low, and here we adopt an extinction of
$A_V = 0.08 \pm 0.02$ from Schlegel et al. (1998).

The distance to IC 1613 was determined from Cepheid variable stars early 
on by Baade (Sandage 1971) and, based in part
on the observations of Cepheids by Freedman (1988), Madore \& Freedman 
(1991) placed IC 1613 at a distance of 765 kpc ($\mu_0 = 24.42 \pm 0.13$).
Saha et al. (1992) observed RR Lyrae variable stars and
derived a distance of 660 kpc ($\mu_0 = 24.10 \pm 0.27$).  Using the
observation of the tip of the red giant branch, Lee et al. (1993)
derived a distance of 714 kpc ($\mu_0 = 24.27 \pm 0.25$), a value
confirmed by the WFPC2 study of an inner field of IC 1613 by
Cole et al. (1999). These values are all consistent within the errors,
but their differences may be dominated by systematic errors.

Cole et al. (1999) presented an initial study of IC 1613 stellar
populations, based on 10700s of integration in both F555W and F814W,
as well as 2600s in F439W, of a field nearer the center of the galaxy.
They found evidence of a continuous star forming history, with the
presence of all expected components of the CMD (main sequence, red
supergiants, blue and red helium burners, asymptotic red giants, red
giants, and a red clump), as well as a hint of a blue horizontal branch.
From the CMD morphology of the central field, Cole et al. derived an
approximate age-metallicity relation for IC 1613, finding it similar in
form to that of the SMC but $\sim 0.3$ dex more metal-poor at any given age.
Their data were not designed for variable star study (the primary goal
of this paper), however, so we will leave a detailed comparison between
their results and those from these data for a future paper on the
stellar populations.

A critical issue in the use of extragalactic stellar distance indicators 
is their sensitivity to changes in metallicity. This question received a 
good deal of attention in the recent literature (e.g., Kennicutt et al.
1998, Sandage, Bell, \& Tripicco 1999, Caputo, and Marconi, \& Musella 2000
for discussions of the Cepheid scale; McNamara 1997, Caputo et al. 2000, and
Demarque et al. 2000 for RR Lyraes; Lee et al. 1993 and Salaris \& Cassisi
1998 for the RGB tip; and Girardi \& Salaris 2000 and Udalski 2000 for the
red clump).
A fundamental point in these discussions is whether offsets due to
metallicity are more or less important than uncertainties in the various 
luminosity zero points. Since the first rung of the variable star distance
ladder is the Magellanic Clouds, which are moderately metal-poor,
observations of similar galaxies, such as IC~1613, provide the opportunity 
to empirically test for the importance of modest variations in metallicity
among metal-poor systems.  In the present observations, we are 
able to measure the magnitudes of four independent ``standard candles'' --
RR Lyrae variable stars, Cepheid variable stars, the tip of the red giant
branch, and the mean magnitude of the red clump -- in a
galaxy that has slightly lower metallicity than the Small Magellanic
Cloud. By comparing these four distance determinations to those of the
Magellanic Clouds, we are able to test for systematic differences
associated with each indicator and potential metallicity dependencies.

\section{Data and Reduction}

\subsection{Observations}
WFPC2 observations of a field in the halo of IC 1613 were obtained during
22-27 August 1999, as part of program GO-7496, whose purpose is to
investigate the stellar populations of dwarf irregular galaxies.  The
field center
(01$^h$ 04$^m$ 26.7$^s$, +02$^{\circ}$ 03' 16'', J2000) is located 11.6'
southwest of the center of the galaxy and 6.7' southwest of the field
studied by Cole et al. (1999).  There is no overlap with the field
examined by Saha et al. (1992) in the previous IC 1613 RR Lyrae study, which
is 14.6' to the west, or with the field of Antonello et al. (1999).
The data consist of 48 1200s
images: 16 in F555W and 32 in F814W, spread evenly among four dithering
pointings.  Each orbit consisted of two images in the same filter to aid in
cosmic ray removal.  A 25th orbit was used to obtain two F656N (H$_\alpha$)
images, which were used for other purposes, but not for the variable star
work that is the topic of this paper.

\subsection{Reductions}
The data were obtained from the STScI archive using on the fly calibration,
and thus were pipeline-calibrated using the best available calibration
images at the time of retrieval.  The images were then reduced using the
HSTphot package (Dolphin 2000a).  The data quality image (c1f) was used to
mask bad and questionable pixels, and the pairs of images from each orbit
were combined for cosmic ray removal, producing 8 clean 2400s images in
F555W and 16 images in F814W.  A deep image produced from combining all
eight F814W images at the first pointing is shown in Figure \ref{figImage}.
The sky image (which contains calculated sky values at each pixel and is
determined before running photometry for uncrowded images for greater
efficiency) was then calculated and hot pixels removed using the HSTphot
utilities.
\placefigure{figImage}

Photometry was made using the \textit{multiphot} routine of HSTphot,
which solves the photometry simultaneously on multiple images (all 24, in
this case) in order to reduce the number of free parameters.  The detection
threshold was a minimum signal to noise of 3.5 in both the combined F555W
and combined F814W measurements.  CTE loss
corrections and calibrations were made using Dolphin (2000b).  Because
of the presence of bright, isolated stars, PSF solutions and aperture
corrections were made for each chip of each image.  The color-magnitude
diagram is shown in Figure \ref{figCMD}, using all stars with the
goodness-of-fit parameter $\chi \le 1.5$, $|$object sharpness$| \le 0.3$
(sharpness of a perfect star is zero), and total signal-to-noise of at
least 5 in both F555W and F814W.  In order to eliminate poor detections,
these requirements were also made on the detections at each epoch, with
detections failing to meet the $\chi$, sharpness, and signal-to-noise
criteria eliminated.  To verify the accuracy of our photometry, we reduced
the data at one pointing independently, using DoPHOT procedures described
by Saha et al. (1996).  This comparison is shown in detail in Dolphin
(2000a), with agreement to within 0.01 magnitudes in both filters.
\placefigure{figCMD}

Before proceeding with the variable star search, it was necessary to
determine the maximum accuracy of the photometry.  This was done
by comparing the magnitudes of well-measured upper red giant branch
stars at each epoch with the combined magnitudes.  This comparison,
which tests both the reliability of the photometry and that of the
aperture corrections, showed a median scatter of 0.015 magnitudes.  This
value was adopted as the minimum error in the variable star work, with
all smaller photometric uncertainties increased to 0.015 magnitudes.
The source of this error is a combination of photometric error from the
undersampled images and error in the aperture corrections.

Figure \ref{figScatter} shows the scatter (individual epoch minus combined
magnitude) for all stars.  The locations of the Cepheids (F555W of
$\sim22-23$) and RR Lyraes (F555W of $\sim25$) are shown by the excess
scatter, and the limiting accuracy of 0.015 magnitudes at the bright end
is clear.  Otherwise, the figure is typical for a star-forming dwarf.
\placefigure{figScatter}

\subsection{Variable Star Identification}

Variable star candidates were identified using a procedure similar to that
described by Saha \& Hoessel (1990).  For a star to be considered a candidate
variable, it had to meet four criteria.  First, the star had to have
good photometry in at least 16 of the 24 epochs, and an rms scatter of
at least 0.08 magnitudes.  Second, the reduced $\chi^2$ of the photometry,
as defined by
\begin{equation}
\chi^2 = \frac{1}{N_{F555W}+N_{F814W}} ( \sum_{i=1}^{N_{F555W}} \frac{(F555W_i - \overline{F555W})^2}{\sigma_i^2} + \sum_{i=1}^{N_{F814W}} \frac{(F814W_i - \overline{F814W})^2}{\sigma_i^2} ),
\end{equation}
had to be at least 6.25, where $N_{F555W}$ and $N_{F814W}$ are the numbers of
exposures in the two filters, and $F555W_i$ and $F814W_i$ are the magnitudes
at each epoch, and $\overline{F555W}$ and $\overline{F814W}$ are the mean
magnitudes for each star.  The minimum value of 6.25 corresponds to mean
deviation of 2.5$\sigma$, and was determined empirically for these data.
A second $\chi^2$ test was made in order to
reduce the ability of single bad points to cause a false detection.  The
one-third of the points contributing the most to $\chi^2$ were eliminated,
and the reduced $\chi^2$ recalculated from the remaining points.  If this
value was not at least 0.25, the star was eliminated from the list of
candidate variables.  Again, the value of 0.25 was determined empirically.

Finally, a modified Lafler-Kinman algorithm (Lafler \& Kinman 1965) was
used to test the stars for periodicity.  This was implemented by computing
$\Theta$ for periods between 0.1 and 5.0 days.  The $\Theta$ parameter is
calculated by determining the light curve for a trial period and using the
equation
\begin{equation}
\Theta(p) = \frac {\sum_{i=1}^{N}(m_i - m_{i+1})^2}{\sum_{i=1}^{N}(m_i - \overline{m})^2},
\end{equation}
where $N$ is the number of exposures for a given filter, $m_i$ is the
magnitude at epoch $i$, and $\overline{m}$ is the mean magnitude.  If
the trial period is incorrect, $m_i - m_{i+1}$ will be the difference
between two random points, $\sqrt{2}$ times the rms scatter, producing
a $\Theta$ of 2.  However, if the trial period is correct, the difference
between adjacent points will scale as 1/$N$, producing a $\Theta$ that
scales as 1/$N^2$.  A goodness of periodicity parameter can then be
defined as $\Lambda(p) = 2/\Theta(p)$.  For this study, the off-period
$\Theta$ was defined to be the 90th percentile value of $\Theta$ over the
range of trial periods, giving our goodness of periodicity parameter of
$\Lambda(p) = \Theta_{90} / \Theta(p)$.

For the present data set, however, a sufficient number of observations was
made in both F555W and F814W for light curve measurements to be made in
both filters.  Thus a combined goodness of periodicity parameter needed to
be developed.  Given that $\sqrt{\Lambda}$ scales as the number of
observations in a given filter when at the correct period, a reasonable
combined parameter would be
\begin{equation}
\Lambda = 0.25 (\sqrt{\Lambda_{F555W}} + \sqrt{\Lambda_{F814W}})^2,
\end{equation}
where the constant of 0.25 is included to force the off-period $\Lambda$
to 1.  In the general case, a goodness of periodicity for any number of
filters can be calculated with
\begin{equation}
\Lambda = (\frac{1}{N_{filt}} \sum_{i=1}^{N_{filt}} \sqrt{\Lambda_i})^2.
\end{equation}
For this study, $\Lambda$ was required to be at least 2.0 for a star to be
considered a candidate variable.

We note that, because $\Theta$ is statistically independent of amplitude
(doubling the amplitude would leave $\Theta$ unchanged, for example), our
determination of $\Lambda$ does not account for the larger amplitudes of RR
Lyraes and Cepheids in $F555W$.  It is not immediately obvious if or how
such an accounting should be made.  Nevertheless, we have experimented with
other algorithms for the calculation of $\Lambda$ and $\Theta$, such as
\begin{equation}
\Theta(p) = \frac {\sum_{i=1}^{N_{F555W}}(F555W_i - F555W_{i+1})^2 + \sum_{i=1}^{N_{F814W}}(F814W_i - F814W_{i+1})^2}{\sum_{i=1}^{N_{F555W}}(F555W_i - \overline{F555W})^2 + \sum_{i=1}^{N_{F814W}}(F814W_i - \overline{F814W})^2},
\end{equation}
and find the selection of variables and their periods to be quite robust,
regardless of the choice of algorithm.

Out of the 12983 total stars, the steps listed above selected 57 variable star
candidates, which were examined interactively.  As this study is primarily
concerned with pulsating variables, stars that were variable in only
one filter and stars whose F555W and F814W light curves were out of phase
were removed from the candidate list, as were false detections, leaving
26 stars in the list.  Eleven fall in the instability
strip above the horizontal branch, and were classified as Cepheids.
Thirteen fall along the horizontal branch, and were classified as RR
Lyraes.  The remaining two are possible eclipsing binaries.  Figure
\ref{figVCMD} shows the IC 1613 CMD, with the variable stars highlighted.
It should be emphasized that our detection efficiency was not 100\%
(and was much lower for stars with period greater than 0.6 days, the
duration of our longest set of consecutive orbits). Thus we cannot rule
out the existence of Population II Cepheids, nor are we confident that
the ``non-variable'' stars falling within the instability strip are
not, in fact, variables that were not detected.  Additionally, while the
CMD from combined photometry is largely complete to $I = 27$, the single
epoch signal-to-noise of stars below the horizontal branch is such that
detection of variables would have been extremely difficult.
\placefigure{figVCMD}

Mean magnitudes were calculated for each variable, using a period-weighted
average
\begin{equation}
\langle m \rangle = -2.5 \log \sum_{i=1}^{N} \frac{\phi_{i+1} - \phi_{i-1}}{2} 10^{-0.4 m_i},
\end{equation}
where $\phi_i$ is the phase and $m_i$ is the magnitude at each point along
the light curve.  These values, still in WFPC2 F555W and F814W magnitudes,
were then transformed to standard $V$ and $I$.

\section{Analysis}

\subsection{RR Lyraes}
Figure \ref{figRR_LC} shows the light curves of the thirteen RR Lyrae
candidates, and Table \ref{tabRR} contains their positions and data.  All
positions are given relative to the F555W images at the first pointing.
The final column in Table \ref{tabRR} lists the quality of the light curve,
from 0 to 4 (although all stars with quality values of 0 or 1 have been
removed). The criteria that are used to determine the light curve quality
are the uniqueness of the period, the presence or lack of bad points, light
curves in phase between the two filters, and the resemblance to a template
light curve of the appropriate class of object. We do not attempt to
distinguish between fundamental-mode and overtone pulsators in the RR
Lyraes, given the relatively poor signal-to-noise at each epoch (the
typical uncertainty in both filters is 0.1 magnitudes). However, we do note
that the short periods of variables V14 and V16 make them likely overtone
pulsators, and that those of V19 and V23 make them possible overtone
pulsators.
\placefigure{figRR_LC}
\placetable{tabRR}

Multiplying the uncertainties in $\langle V \rangle$ by 4 divided by the
light curve quality, and taking an average weighted by $1/\sigma^2$, the
best $\langle V \rangle$ for the sample of 13 RR Lyraes is $25.00 \pm 0.03$
magnitudes.  Eliminating the two possible outliers fainter than V=25.2,
the weighted average is $24.98 \pm 0.03$.  Similarly eliminating the
two possible outliers brighter than V=24.8, the weighted average becomes
$25.02 \pm 0.03$.  Given the very small shift in the mean magnitude
after eliminating the lowest and highest points, it seems reasonable
to adopt the value of $\langle V \rangle = 25.00$ for the mean RR Lyrae
magnitude, while adding the $\pm0.02$ shift in quadrature to the 0.03
magnitude uncertainty, producing a final uncertainty of 0.04 magnitudes.
This mean $V$ magnitude is consistent with the value of
$\langle g \rangle = 24.90 \pm 0.10$ obtained by Saha et al. (1992), which
corresponds to a mean $V$ magnitude of $24.94 \pm 0.10$ (Saha \& Hoessel
1987).  An extinction $A_V$ of 0.08 magnitudes is adopted from Schlegel
et al. (1998) (assuming $A_V/A_B = 3.1/4.1$), with the uncertainty
estimated to be $\pm 0.02$ magnitudes, producing an extinction-corrected
mean $V$ magnitude of $\langle V \rangle_0 = 24.92 \pm 0.04$.

Once the mean $V$ magnitude is established, obtaining a distance estimate
requires a value for the absolute magnitude.  This is generally done by
adopting a value for the mean RR Lyrae metallicity and a preferred
$M_V$ vs. $\logz$ relation.  We can measure the mean RR Lyrae metallicity
via the red giant branch, whose tip spans the color range
$1.45 < (V-I) < 1.62$.  Assuming that RR Lyraes are only produced in
populations older than 10 Gyr, we find an allowable metallicity range of
$-1.5 < \logz < -1.1$ (with $\logz = -1.5$ and $t = 15$ Gyr falling along
the blue edge of our RGB and $\logz = -1.1$ and $t = 10$ Gyr falling along
the red edge), based on isochrones interpolated from those of
Girardi et al. (2000).  Thus we find a value of $\logz = -1.3 \pm 0.2$
for the RR Lyrae metallicity.  For comparison, metallicity calculations
using the Da Costa \& Armandroff (1990) $(V-I)_{M_I=-3.0}$ and Lee et al.
(1993) $(V-I)_{M_I=-3.5}$ calibration produce the identical metallicity
($\logz = -1.3 \pm 0.2$), after conversion to the Carretta \& Gratton
(1997) metallicity scale.

The range of resulting absolute magnitudes is rather large, however,
producing significant uncertainty in the RR Lyrae distance.  Theoretical
models of Girardi et al. (2000) predict an absolute magnitude of
$M_V = 0.60 \pm 0.03$; those of Demarque et al. (2000) produce an absolute
magnitude of $M_V = 0.62 \pm 0.05$; and those of Caputo et al. (2000)
produce an absolute magnitude of $M_V = 0.57 \pm 0.12$.  Carretta et al.
(2000) find a relation based on LMC distances and HIPPARCOS parallax
measurements of subdwarfs that produces an absolute magnitude of
$M_V = 0.59 \pm 0.07$.  Other theoretical models produce brighter RR
Lyraes, with Caloi et al. (1997) and Cassisi et al. (1999) finding
$M_V \approx 0.55$ for the ZAHB; the mean RR Lyrae magnitude being about
0.1 magnitudes brighter.  Observational determinations based on
HIPPARCOS show a larger spread in values, such as those of Fernley et al.
1998, which produces an absolute magnitude of $Mv = 0.73 \pm 0.14$, and
Gratton et al. 1997, which produces an absolute magnitude of
$Mv = 0.45 \pm 0.04$.  We therefore conservatively adopt a value of
$M_V = 0.60 \pm 0.15$, which produces an distance modulus of
$24.32 \pm 0.16$.  We will return to the issue of the RR Lyrae zero point
later, using independent distance measurements to derive the absolute
magnitude of RR Lyraes in IC 1613.

\subsection{Cepheids}
Although the observations were designed primarily to detect RR Lyraes,
eleven short-period Cepheids were also discovered.  Figure
\ref{figCeph_LC} shows their light curves, and Table \ref{tabCeph}
contains their positions and data.  As with Table \ref{tabRR}, the quality
rating ranges from 0 to 4, with 4 being the cleanest light curve and
best-defined period.  The Cepheids were also classified into
fundamental-mode and overtone pulsators, with all but V13 classified
based on their light curves.  As shown by Mantegazza \& Poretti (1992),
overtone pulsators have a Fourier spectrum with a weaker second order than
do fundamental pulsators, meaning that their light curves will be more
sinusoidal.  Ten of the eleven Cepheids were thus classified, and the
period-luminosity relation in Figure \ref{figCeph_PL} shows this
discrimination to be successful, with all overtone pulsators falling well
above the mean period-luminosity relation.  The period of V13 was
poorly-constrained
due to the poor sampling of epochs, and thus it could not be classified
similarly.  However, its position on the period-luminosity relation was
clearly in the space occupied by overtone pulsators, allowing its
classification based on period and magnitude.
\placefigure{figCeph_LC}
\placetable{tabCeph}

The presence of five fundamental-mode Cepheids allows for a second
distance measurement from these data.  One can do this following Madore
\& Freedman (1991),
determining the reddening-free magnitude $W=2.43I - 1.43 V$ (assuming
$A_V/A_I = 1.7$) for the five Cepheids and calculate the reddening-free
absolute magnitude $M_W = -3.049 \log P - 2.40$ using the Madore \&
Freedman (1991) calibration.  Applying this technique to the IC 1613
fundamental Cepheids and using a weighted average of the uncertainties
gives a reddening-corrected distance modulus of $\mu_0 = 24.50 \pm 0.11$
and a mean extinction $A_V$ of $0.16 \pm 0.11$ magnitudes.  The $V$ and
$I$ period-luminosity relations are shown in Figure \ref{figCeph_PL},
with the Cepheids found by Freedman (1988) plotted as well.  As can be
seen in the Figure, the present data are consistent with and provide an
extension to the longer-period data of Freedman (1988).
\placefigure{figCeph_PL}

However, the EROS result (Bauer et al. 1999) of a steepening in the
period-luminosity
relation for periods shorter than two days implies that three of our five
fundamental-mode Cepheids should be eliminated from this calculation.
(This break is shown with the dashed line in Figure \ref{figCeph_PL}.)
Thus the reddening-corrected distance modulus becomes
$\mu_0 = 24.55 \pm 0.18$ for the two remaining fundamental-mode Cepheids.
The large uncertainty is the result of the use of the reddening-free
distance, which multiplies $V$ uncertainties by 2.43 and $I$ by 1.43 and
adds them in quadrature.  Given the low extinction to IC 1613 and the
presence of a good extinction estimate of $A_V = 0.08 \pm 0.02$, the
uncertainty can be lowered by correcting the individual $V$ and $I$
distance moduli for extinction and combining them.

Finally, it should be noted that the $I$ photometry of V3 do not
adequately sample the full range of the light curve, as the earliest
epoch was at a phase of 0.2 after the peak.  Thus the phase-averaged
$\langle I \rangle$ magnitude is biased towards fainter magnitudes.
Because this was not a problem in the $V$ photometry, the procedure of
Labhardt, Sandage, \& Tammann (1997) was used to calculate the correct
mean $\langle I \rangle$, brightening the value from 21.21 to 21.15.  The
data for V6 have a similar problem, but a similar correction is impossible
because neither the $V$ nor $I$ photometry span the entire range of
magnitudes.  However, because the $V$ light curve omits as much of the
trough as the $I$ light curve omits of the peak, these errors should
largely cancel when averaging the $V$ and $I$ distances obtained for V6.

Table \ref{tabCephDist} shows the mean $V$ and $I$ magnitudes corrected
for extinction values
$A_V = 0.08 \pm 0.02$ and $A_I = 0.05 \pm 0.02$.  Absolute magnitudes
$M_V$ and $M_I$ are calculated using Madore \& Freedman (1991), with
their rms scatter adopted as the uncertainties.  Given the smaller
intrinsic scatter in the $I$ period-luminosity relation, the distance for
each of the Cepheids is weighted twice as much in $I$ as in $V$.  Averaging
the values for the two stars gives a best Cepheid distance from these
data of $\mu_0 = 24.45 \pm 0.15$ magnitudes, consistent with the value
of $\mu_0 = 24.42 \pm 0.13$ determined by Madore \& Freedman (1991) from
the Cepheid calibration used here, as well as with previous measurements
of Sandage (1971); McAlary, Madore, \& Davis (1984); and Freedman (1988).
The values in Figure \ref{figCeph_PL} are for this distance and the adopted
extinction, with the dashed line showing the expected $V$ relation for
$P < 2^d$ based on the EROS result.  Because of their larger sample size,
we will adopt the Madore \& Freedman (1991) value in our discussion,
after converting it to our adopted extinction of $A_V = 0.08$ for the
sake of comparison with other distance measurements.  This decreases their
distance to $\mu_0 = 24.40 \pm 0.13$.  We will discuss possible
metallicity effects on the Cepheid distance scale below.
\placetable{tabCephDist}

\subsection{Other Distance Measurements}
Although variable stars provide excellent distance indicators, there
remain questions about the zero-point calibrations at the 20\% level,
as well as uncertainties regarding their dependencies on the metallicity
of the parent population of stars.  These issues will be discussed in the
following section.  IC 1613 provides us also with additional distance
benchmarks, which provide further constraints for studying this problem.

The red giant branch (RGB) tip (Lee, Madore, \& Freedman 1993) and red
clump (Paczynski \& Stanek 1998, Udalski 2000) provide two possible
standard candles.  The RGB tip
in I is especially attractive, given its insensitivity to age and
metallicity in theoretical models (Girardi et al. 2000, for example).
From the CMD in Figure \ref{figCMD} and an edge-detection algorithm, we
measure an RGB tip at $I = 20.40 \pm 0.09$.  However, this value is something
of an upper limit to the RGB tip magnitude, as the small number of stars
present in the upper RGB makes it plausible that the theoretical RGB
extends above the position of these stars.  With an upper RGB luminosity
function of roughly 7 stars per 0.1 magnitudes, our 68\% confidence
limit for the true RGB tip is 0.03 magnitudes above the brightest star,
increasing our uncertainty in the RGB tip measurement to $0.10$ magnitudes.

We can derive a more accurate RGB tip by use of the denser field studied
by Cole et al. (1999), which contains significantly more red giants.
Reducing those data with HSTphot and applying the Dolphin (2000b) CTE
correction and calibration, we find the RGB tip at $I = 20.35 \pm 0.07$,
consistent with that determined for the present data (the 0.07 magnitude
uncertainty a result of possible RGB-AGB confusion near the tip).
The luminosity
function for the upper RGB is shown in Figure \ref{figTRGB}.  Applying our
adopted extinction correction of $A_I = 0.05 \pm 0.02$ produces an
extinction-corrected RGB tip magnitude of $I_0 = 20.30 \pm 0.07$.  We find
the photometry of Cole et al. nearly 0.1 magnitudes brighter, and note
that we adopt the HSTphot-based photometry for the inner field, as it
is more consistent with the photometry for the outer field (which was
consistent with DoPHOT reductions of the same data), and because
of the availability of an HSTphot-based calibration and CTE correction
(Dolphin 2000b).
\placefigure{figTRGB}

The absolute magnitude at the RGB tip can be estimated either empirically
(Lee et al. 1993) or theoretically (Girardi et al. 2000). Empirical
calibrations rest on the globular cluster distance scale, which is itself
dependent upon an assumed RR Lyrae absolute magnitude calibration, and
thus cannot be used to make an \textit{independent} distance measurement
for our study.  Theoretical techniques, on the other hand, are subject to
uncertainties in the input physics (Castellani \& degl'Innocenti 1999) and
the transformation from physical parameters (luminosity and temperature) to
observed magnitudes and colors. In order to produce an independent
distance measurement and to account for the uncertainties, we thus adopt
an absolute magnitude calculated from interpolated isochrones from Girardi
et al. (2000), while conservatively adding the difference between this
value and that calculated from the Lee et al. (1993) calibration to our
uncertainties. Fitting the isochrones to our red giant branch for ages from
2 to 15 Gyr, we measure
a mean metallicity of $\logz = -1.15 \pm 0.2$ and absolute magnitude of
$M_I = -4.02 \pm 0.02$.  For comparison, the calibration given by Lee et al.
(1993), using the equivalent metallicity on the Zinn \& West (1984) scale
($\logz = -1.37$ from the conversion given by Carretta \& Gratton 1997),
produces an absolute magnitude of $M_I = -4.06 \pm 0.04$, consistent but
0.04 magnitudes brighter.  We thus arrive at an RGB tip absolute magnitude of
$M_I = -4.02 \pm 0.05$ and a distance modulus of $\mu_0 = 24.32 \pm 0.09$.

We note that the RGB tip calibration of Salaris \& Cassisi (1998) provides
a significantly more luminous tip, at $M_I = -4.25 \pm 0.02$, which would
increase the IC 1613 distance modulus to $\mu_0 = 24.55 \pm 0.07$.  We
are unable to reconcile this value with the red clump distance (below),
regardless of the star formation history assumed for IC 1613.  Because the
red clump is the only one of the four distance indicators calibrated
directly by HIPPARCOS, we choose to adopt the calibration above of
$M_I = -4.02 \pm 0.05$, with the RGB tip distance modulus of
$\mu_0 = 24.32 \pm 0.09$.

Another independent measurement of the distance is provided by the mean $I$
magnitude of the red clump. Although $M_I$(RC) varies significantly with
both age and metallicity (for example, Girardi et al. 1998 and Girardi \&
Salaris 2000 for theoretical work; Sarajedini 1999 for empirical work),
Girardi \& Salaris (2000) have demonstrated that it can be adequately
modeled, and used as an accurate distance indicator that provides distances
consistent with other, more-tested, distance measurements.

The measurement of the red clump mean magnitude is relatively easy, aided
by the separation of the two features in the CMD.  Figure \ref{figILF}
shows the $I$ luminosity function for stars with $0.8 \le V-I \le 0.95$ and
$22 \le I \le 25.5$, with the best-fit polynomial (RGB) plus
Gaussian (red clump) shown.  The peak of the Gaussian falls at
$I = 23.90 \pm 0.01$ (uncertainties derived using bootstrap tests and
a variety of bin sizes), which corresponds to an extinction-corrected value
of $I_0 = 23.85 \pm 0.02$.  For comparison, our HSTphot-reduced photometry
of the inner field (the field studied by Cole et al. 1999) produces a
slightly brighter red clump magnitude of $I = 23.86 \pm 0.01$.  The
difference of 0.04 magnitudes is statistically significant; the most likely
explanation is the presence of a larger fraction of young ($t \lesssim 1$
Gyr) stars in the inner field.
\placefigure{figILF}

Determination of the red clump absolute magnitude, however, is significantly
more difficult.  Udalski (2000) provides an empirical calibration of
$M_I(RC)$ vs. metallicity, but this calibration would have to be
extrapolated by nearly a dex in metallicity for application to IC 1613
and is based on the assumption that there is no age-metallicity relation
in the Galactic disk.  Neither of those problems can be easily solved
analytically, so we instead employ the Girardi et al. (2000) theoretical
isochrones as an accurate relative absolute magnitude indicator, with
which we can compare the local disk clump with that of IC 1613 (a method
very similar to ``method 2'' of Girardi \& Salaris 2000).  We produced a
synthetic
CMD for the local Galactic disk based on the age-metallicity relation and
star ages of Table 3 of Rocha-Pinto et al. (2000), and a set of 15 synthetic
CMDs for IC 1613 based on a range of possible age-metallicity relations
and star formation histories consistent with the results of Cole et al.
(1999).  Applying the usual Gaussian clump plus quadratic RGB fit to the
synthetic CMDs, we measure a difference of $\delta M_I = -0.22 \pm 0.08$
magnitudes between the IC 1613 red clump and the Galactic disk red clump,
in the sense that the IC 1613 clump is brighter.  Combined with the
HIPPARCOS-based local red clump calibration of $M_I = -0.23 \pm 0.02$
(Stanek \& Garnavich 1998), we derive a semi-empirical IC 1613 red clump
absolute magnitude of $M_I = -0.45 \pm 0.09$ and a red clump distance
modulus of $\mu_0 = 24.30 \pm 0.09$.

Thus, two additional independent distances can be obtained by use of the
RGB tip and the red clump.  The RGB tip distance is corrected for age and
metallicity relatively easily, but difficulty in observational determination
of the tip magnitude and uncertainty in the calibration create additional
error.  In contrast, the red clump position is easily measured with high
precision, but the systematics from age- and metallicity-dependencies are
not as well-constrained.

\section{Discussion}

\subsection{Relative IC 1613 -- SMC -- LMC Distances}

With the availability of four independent distance measurements to
IC 1613, we can attempt to determine relative distances between IC 1613
and the Magellanic Clouds.  As the zero point subtracts out when measuring
relative distances, this allows us, for the time being, to ignore the
uncertainties in the calibrations.  Table \ref{tabSMCcomp} and Figure
\ref{figDists} show these comparison values.
\placetable{tabSMCcomp}
\placefigure{figDists}

For RR Lyraes, we adopt an absolute magnitude of $M_V = 0.60$ at
$\logz = -1.3$.  We conservatively adopt a metallicity dependence of
$0.25 \pm 0.10$ magnitudes per dex, which is consistent with both the
steep scale ($\frac{dM_V}{d\logz} \sim 0.3$; for example, McNamara 1997)
and the shallow scale ($\frac{dM_V}{d\logz} \sim 0.2$; for example,
Carretta et al. 2000), producing absolute magnitudes of
$M_V = 0.58 \pm 0.01$ at $\logz = -1.4$ (NGC 121 in the SMC) and
$M_V = 0.45 \pm 0.06$ at $\logz = -1.9$ (the clusters used for the LMC
measurement).  The resulting relative distances are shown in the first
row of Table \ref{tabSMCcomp}; we note that absolute distances would
need to also include the zero point error of $\pm 0.15$ magnitudes.

The Cepheid distance scale is based on the Madore \& Freedman (1991)
calibration, which assumes an LMC distance modulus of 18.50.  This
calibration does not include any metallicity dependence, a potential
source of error given the significant metallicity range covered by
these three objects (from $\logz \simeq -1.0$ for IC 1613 to
$\logz \simeq -0.4$ for the LMC).  As literature values for the dependence
are more varied than are those for the RR Lyrae dependence, we will not
attempt to make a correction here; instead the values in the second row
of Table \ref{tabSMCcomp} assume no metallicity dependence.

Rows three and four of Table \ref{tabSMCcomp} similarly give RGB tip and
red clump distances to the three objects.  We adopt the Girardi et al.
(2000) models to provide the absolute magnitudes of the RGB tips, and
the Girardi \& Salaris semi-empirical calibrations of the Magellanic
Cloud red clump absolute magnitudes.

We note that both the relative IC 1613 $-$ SMC distances and the relative
IC 1613 $-$ LMC distances are consistent between the four measurement
methods.  Using a weighted average of the four (by $\frac{1}{\sigma^2}$)
relative IC 1613 $-$ SMC distance measurements, we find a value of
$5.44 \pm 0.05$, corresponding to a linear distance ratio of
$\frac{d_{IC 1613}}{d_{SMC}} = 12.2 \pm 0.3$.  Because of the possible
metallicity dependence in the
Cepheids and the large metallicity difference between IC 1613 and the
LMC, we average only the RR Lyrae, RGB tip, and red clump relative
distances, finding a relative IC 1613 $-$ LMC distance of $5.83 \pm 0.06$
(a linear distance ratio of $\frac{d_{IC 1613}}{d_{LMC}} = 14.7 \pm 0.4$).

\subsection{Metallicity Dependencies of Distance Indicators}

Although the uncertainties are significant, we also wish to use the
metallicity baseline in this comparison to examine the effects of
metallicity on the RR Lyrae and Cepheid distance measurements.  We first
note that, since the four relative IC 1613 $-$ LMC distance measurements
are all consistent, our data are consistent with the metallicity
dependencies adopted in the previous section:
$\frac{d M_V}{d\logz} = 0.25$ for RR Lyraes and
$\frac{d M_V}{d\logz} = \frac{d M_I}{d\logz} = 0$ for Cepheids.

We first address the issue of the RR Lyrae metallicity dependence.
Combining the red clump and RGB tip distances (which have both been
corrected for population effects) to determine relative distances to
the three objects, we derive reddening-corrected absolute magnitudes of
$M_V (IC 1613) = 0.61 \pm 0.08$, $M_V (SMC) = 0.58 \pm 0.08$, and
$M_V (LMC) = 0.41 \pm 0.10$.  Fitting these values to a straight line
(and adopting the metallicities from the previous section) with a
least-$\chi^2$ algorithm, we find a
metallicity dependence of $\frac{d M_V}{d\logz} = 0.34 \pm 0.20$ magnitudes
per dex for the RR Lyrae absolute magnitude, a large value but
consistent with more robust estimates of the metallicity dependency of the
RR Lyrae absolute magnitudes (for example, Sandage et al. 1999, McNamara
1997, Layden et al. 1996, Carney, Storm, \& Jones 1992, Sandage \&
Cacciari 1990, and Liu \& Janes 1990).  As our measured value has a large
uncertainty, we will continue to use the adopted dependence of
$\frac{d M_V}{d\logz} = 0.25 \pm 0.10$ magnitudes per dex.

For the Cepheids, we adopt recent metallicities of
$\logz \simeq -1.0$ for IC 1613 (based on an isochrone fit to
these data), $\logz \simeq -0.8$ for the SMC (based on the
cluster age-metallicity relation given by Olszewski, Suntzeff, \& Mateo
1996), and $\logz \simeq -0.4$ for the LMC (also based on the
cluster age-metallicity relation given by Olszewski et al. 1996).
Following our procedure from the RR Lyrae dependence measurement, we will
compare the Cepheid distances to those from the other distances (this
time using the RR Lyrae, red clump, and RGB tip distances to measure the
``true'' distances).  From the data in Table \ref{tabSMCcomp}, we measure
Cepheid $-$ ``true'' distances of $0.08 \pm 0.14$ for IC 1613,
$0.03 \pm 0.06$ for the SMC, and $0.01 \pm 0.05$ for the LMC.  A
least-$\chi^2$ fit to these points produces a metallicity dependence of
$-0.07 \pm 0.16$ magnitudes per dex in the Cepheid distances, consistent
with zero or with the small metallicity dependencies determined empirically
by Kennicutt et al. (1998) and theoretically by Sandage, Bell, \& Tripicco
(1999) and Alibert et al. (1999).  Despite the large uncertainty, these
data appear to rule out extreme values of the metallicity dependence,
such as those of Caputo, Marconi, \& Musella (2000), Beaulieu et al. (1997),
and Gould (1994).  We will adopt a conservative correction of
$-0.1 \pm 0.2$ magnitudes per dex, and correct the IC 1613 Cepheid distance
(taken from Madore \& Freedman 1991) to $\mu_0 = 24.34 \pm 0.18$.
This value is still based on an assumed LMC distance modulus of 18.50, and
adding an additional 0.1 magnitudes of uncertainty in that value produces
our best IC 1613 Cepheid distance of $\mu_0 = 24.34 \pm 0.20$.

\subsection{The Distance to IC 1613 and RR Lyrae Calibration}

After applying the corrections above, we have four distance measurements
to IC 1613, summarized in Table \ref{tabDist}.  Taking a weighted average
(again weighting by $\frac{1}{\sigma^2}$), we measure the IC 1613
distance modulus to be $\mu_0 = 24.31 \pm 0.06$, corresponding to a
distance of $730 \pm 20$ kpc.  Although we have, in a sense, required the
four distance measurements to be consistent in the previous section and
thus have the possibility of circularity, we note that the RR Lyrae, RGB
tip, and red clump distances are taken from section 3.3 (the RR Lyrae
metallicity dependence measured above does not factor into the IC 1613
distance).  Additionally, our weighted average of the distance measurements
is $24.31 \pm 0.06$, whether or not the Cepheid distance is included
(because of its high uncertainty).  For completeness, we note that, had
we used the Burstein \& Heiles (1982) extinction maps instead of the
Schlegel et al. (1998) maps, we would have used extinctions of $A_V = 0.02$
and $A_I = 0.01$, and arrived at a distance modulus of
$\mu_0 = 24.36 \pm 0.06$.
\placetable{tabDist}

We note that, given the accurate RGB tip and red clump measurements, we
are able to work ``backwards'' to determine the RR Lyrae absolute magnitude.
Removing the RR Lyrae distance from the weighted average, we arrive at an
IC 1613 distance of $\mu_0 = 24.31 \pm 0.07$.  Combining this with our
reddening-corrected mean magnitude of $V_0 = 24.92 \pm 0.04$, we calculate
the absolute magnitude of IC 1613 RR Lyraes to be $0.61 \pm 0.08$.

\section{Summary}

We have presented photometry, variable star analysis, and a series of
distance measurements of a WFPC2 field in the halo of IC 1613.  We found
thirteen RR Lyrae stars and a mean extinction-corrected magnitude of
$\langle V \rangle_0(RR) = 24.92 \pm 0.04$.  The presence of these stars
confirms the existence of an old horizontal branch, consistent with the
ground-based results of Saha et al. (1992).  We also found eleven
short-period Cepheids, two of which were fundamental mode with sufficiently
long periods to determine a distance.  Finally, we applied RGB tip and red
clump distance measurements to IC 1613, determining distances for each.
The summary of our values is given in Table \ref{tabDist}, along with the
primary sources of error in those four measurement.

We assume that the RGB tip distance is the most robust of the four, given
that the dependencies on age and metallicity are very small and have been
calibrated.  However, the small field of view of WFPC2 limits our ability to
accurately measure the position of the tip, limiting the accuracy of our
measurement to $\mu_0(RGB) = 24.32 \pm 0.09$.  The red clump distance,
on the other hand, has significant calibration uncertainty based on the
age dependence, but its position can be accurately measured in these data.
Adding the population-dependencies into our uncertainties, we find a red
clump distance of $\mu_0(RC) = 24.30 \pm 0.09$.

Our sample of Cepheids was insufficient to produce an accurate distance
measurement, but we were able to confirm that our two $\sim3$ day
fundamental-mode Cepheids were consistent with the Cepheid distance
obtained by Madore \& Freedman (1991).  We were also able to estimate,
via comparisons with LMC and SMC distances, a metallicity dependence of
$-0.07 \pm 0.16$ magnitudes per dex.  Applying a conservative estimate of
the metallicity dependence ($-0.1 \pm 0.2$ magnitudes per dex) and an
extinction of $A_V = 0.08$ from Schlegel et al. (1998) to the Madore \&
Freedman (1991) produces a corrected IC 1613 Cepheid distance modulus of
$\mu_0(Ceph) = 24.34 \pm 0.20$.  Combining this with our other distance
measurements with a weighted average, we arrive at our best IC 1613
distance modulus of $\mu_0 = 24.31 \pm 0.06$, corresponding to a
distance of $730 \pm 20$ kpc.

A similar treatment was given to the RR Lyrae, producing a metallicity
dependence of $0.34 \pm 0.20$ magnitudes per dex in the $V$ absolute
magnitude, consistent with literature values.  Given the wide variety
of RR Lyrae absolute magnitudes in the literature, we also found it
useful to measure the IC 1613 RR Lyrae absolute magnitude, given our
observed mean $V$ magnitude and the distances calculated through other
measurements.  We calculated a mean $M_V$ of $0.61 \pm 0.08$ at
$\logz \simeq -1.3$, a value consistent with both the ``faint''
calibration of Fernley et al. (1998) and the ``bright'' calibration of
Gratton et al. (1997).

We note that, when each distance measurement is properly calibrated and
corrected for population effects, the RR Lyrae, Cepheid, RGB tip, and red
clump distance techniques produce consistent relative distances between
IC 1613 and the Magellanic Clouds.  We find a relative IC 1613 $-$ SMC
distance modulus of $5.44 \pm 0.05$ and a relative IC 1613 $-$ LMC
distance modulus of $5.83 \pm 0.06$.  We also note that all four
distance indicators produce consistent distances to IC 1613.

\acknowledgments

Support for this work was provided by NASA through grant numbers
GO-07496 and GO-02227.06-A from the Space Telescope Science Institute,
which is operated by AURA, Inc., under NASA contract NAS 5-26555.
EDS is grateful for partial support from NASA LTSARP grants No. NAGW-3189
and NAG5-9221.

\clearpage
\begin{figure}
%\plotone{figs/fig01.ps}
\caption{F814W WFPC2 image of IC 1613 field, combined from 8 1200s exposures \label{figImage}}
\end{figure}

\begin{figure}
%\plotone{figs/fig02.eps}
\caption{$(V-I)$, $I$ Color-magnitude diagram (12983 stars), calculated from mean magnitudes in all epochs. Poorly-fit stars ($\chi > 1.5$ or $|$sharpness$| > 0.3$) are not included. Error bars show typical $(V-I)$ uncertainties as a function of $I$. Absolute magnitudes (on the y-axis on the right) are calculated assuming $I - M_I = 24.36$. \label{figCMD}}
\end{figure}

\begin{figure}
%\plotone{figs/fig03.eps}
\caption{Photometry accuracy for the IC 1613 data.  Values given are magnitudes at individual epochs minus mean magnitudes.  The locations of the Cepheids (F555W of $\sim22-23$) and RR Lyraes (F555W of $\sim25$) are shown by the excess scatter, and the limiting accuracy of 0.015 magnitudes at the bright end is clear. \label{figScatter}}
\end{figure}

\begin{figure}
%\plotone{figs/fig04.eps}
\caption{CMD of IC 1613 field, showing the periodic variable stars. Circles are the Cepheids, diamonds the RR Lyraes, and triangles the two possible eclipsing binaries. The detection efficiency was not 100\%, and thus there are many stars in the instability strip. Absolute magnitudes (on the y-axis on the right) are calculated assuming $I - M_I = 24.36$. \label{figVCMD}}
\end{figure}

\begin{figure}
%\plotone{figs/fig05.ps}
\caption{Light curves of 13 candidate RR Lyraes. \label{figRR_LC}}
\end{figure}

\begin{figure}
%\plotone{figs/fig06.ps}
\caption{Light curves of 11 candidate Cepheids. \label{figCeph_LC}}
\end{figure}

\begin{figure}
%\plotone{figs/fig07.eps}
\caption{Cepheid period-luminosity relation.  Solid circles are fundamental pulsators, open circles are overtone pulsators, and crosses are the Cepheids found by Freedman (1988). All but one of the Cepheids were classified by light curve shape. The solid lines are mean P-L relations from Madore \& Freedman (1991), with a distance modulus of 24.45 and extinction of $A_V = 0.08$. The dashed line is the break at P=2 days, based on Bauer et al. (1999) \label{figCeph_PL}}
\end{figure}

\begin{figure}
%\plotone{figs/fig08.eps}
\caption{$I$-band luminosity function along the RGB. Our measured TRGB is at $I = 20.35 \pm 0.07$, and is marked by the vertical line at that position. \label{figTRGB}}
\end{figure}

\begin{figure}
%\plotone{figs/fig09.eps}
\caption{$I$-band luminosity function for stars with $0.8 \le V-I \le 0.95$ and $22 \le I \le 25.5$.  The line is the best fit, using a quadratic polynomial to fit the RGB and a Gaussian to fit the red clump. \label{figILF}}
\end{figure}

\begin{figure}
%\plotone{figs/fig10.eps}
\caption{Distance modulus differences between IC 1613 and the Magellanic Clouds, using four distance measurements. The top panel shows the IC 1613 - SMC differences and the bottom panel the IC 1613 - LMC differences. We note that all four distance measurement techniques produce consistent distance ratios. \label{figDists}}
\end{figure}

\clearpage
\begin{deluxetable}{lcrrrrcc}
\tablecaption{RR Lyraes \label{tabRR}}
\tablehead{
\colhead{ID} &
\colhead{chip} &
\colhead{X} &
\colhead{Y} &
\colhead{$\langle V \rangle$} &
\colhead{$\langle V-I \rangle$} &
\colhead{period} &
\colhead{Q}}
\startdata
V5 & WFC2 & 174.01 & 582.68 & $25.222\pm0.087$ & $0.678\pm0.097$ & 0.59 & 3 \nl
V8 & WFC2 & 498.73 & 101.37 & $24.948\pm0.039$ & $0.333\pm0.051$ & 0.50 & 2 \nl
V14 & WFC3 & 368.24 & 153.10 & $25.098\pm0.083$ & $0.345\pm0.092$ & 0.31 & 3 \nl
V15 & WFC3 & 447.66 & 154.61 & $25.077\pm0.056$ & $0.554\pm0.067$ & 0.63 & 3 \nl
V16 & WFC3 & 519.91 & 703.76 & $25.097\pm0.068$ & $0.463\pm0.074$ & 0.34 & 2 \nl
V17 & WFC3 & 559.79 & 569.96 & $24.790\pm0.102$ & $0.410\pm0.111$ & 0.62 & 3 \nl
V18 & WFC3 & 599.32 & 495.63 & $24.971\pm0.085$ & $0.501\pm0.090$ & 0.65 & 4 \nl
V19 & WFC3 & 661.68 & 400.55 & $25.233\pm0.152$ & $0.368\pm0.170$ & 0.43 & 4 \nl
V20 & WFC3 & 770.59 & 790.22 & $24.728\pm0.135$ & $0.287\pm0.142$ & 0.61 & 2 \nl
V21 & WFC4 & 343.38 & 584.32 & $24.987\pm0.120$ & $0.529\pm0.132$ & 0.60 & 4 \nl
V22 & WFC4 & 585.38 & 573.20 & $25.058\pm0.091$ & $0.550\pm0.107$ & 0.58 & 2 \nl
V23 & WFC4 & 610.83 & 108.18 & $24.843\pm0.112$ & $0.406\pm0.121$ & 0.39 & 4 \nl
V24 & WFC4 & 691.07 & 215.75 & $24.890\pm0.053$ & $0.707\pm0.094$ & 0.48 & 2 \nl
\enddata
\end{deluxetable}

\clearpage
\begin{deluxetable}{lcrrrrccc}
\tablecaption{Cepheids \label{tabCeph}}
\tablehead{
\colhead{ID} &
\colhead{chip} &
\colhead{X} &
\colhead{Y} &
\colhead{$\langle V \rangle$} &
\colhead{$\langle I \rangle$} &
\colhead{period} &
\colhead{mode} &
\colhead{Q}}
\startdata
V1 & WFC1 & 290.70 & 582.93 & $22.103\pm0.064$ & $21.650\pm0.049$ & $1.67\pm0.17$ & OT & 2 \nl
V2 & WFC2 & 120.54 & 360.19 & $22.584\pm0.059$ & $21.990\pm0.024$ & $1.31\pm0.05$ & OT & 4 \nl
V3 & WFC2 & 133.33 & 429.22 & $21.802\pm0.120$ & $21.211\pm0.060$ & $3.31\pm0.06$ & FM & 4 \nl
V4 & WFC2 & 146.33 & 595.41 & $22.606\pm0.074$ & $22.040\pm0.051$ & $1.09\pm0.08$ & OT & 3 \nl
V6 & WFC2 & 387.65 & 282.16 & $21.548\pm0.162$ & $21.083\pm0.090$ & $3.03\pm0.05$ & FM & 4 \nl
V7 & WFC2 & 482.48 & 424.66 & $23.006\pm0.093$ & $22.455\pm0.065$ & $1.34\pm0.06$ & FM & 4 \nl
V9 & WFC2 & 505.88 &  94.89 & $22.675\pm0.110$ & $22.095\pm0.076$ & $1.75\pm0.06$ & FM & 4 \nl
V10 & WFC2 & 552.35 &  83.57 & $23.144\pm0.150$ & $22.627\pm0.071$ & $1.05\pm0.03$ & FM & 3 \nl
V13 & WFC3 & 366.95 & 129.45 & $21.403\pm0.045$ & $20.760\pm0.030$ & $2.82\pm1.20$ & OT & 3 \nl
V25 & WFC4 & 716.29 & 460.78 & $22.565\pm0.147$ & $22.206\pm0.051$ & $0.95\pm0.05$ & OT & 4 \nl
V26 & WFC4 & 717.22 & 589.95 & $22.769\pm0.050$ & $22.211\pm0.032$ & $1.00\pm0.07$ & OT & 3 \nl
\enddata
\end{deluxetable}

\clearpage
\begin{deluxetable}{lrrrrrr}
\tablecaption{Cepheid Distances \label{tabCephDist}}
\tablehead{
\colhead{ID} &
\colhead{$\langle V \rangle_0$} &
\colhead{$\langle I \rangle_0$} &
\colhead{period} &
\colhead{$M_V$} &
\colhead{$M_I$} &
\colhead{$\mu_0$}}
\startdata
V3 & $21.80\pm0.12$ & $21.15\pm0.06$ & $3.31\pm0.06$ & $-2.83\pm0.27$ & $-3.40\pm0.18$ & $24.58\pm0.21$ \nl
V6 & $21.55\pm0.16$ & $21.08\pm0.09$ & $3.03\pm0.05$ & $-2.73\pm0.27$ & $-3.28\pm0.18$ & $24.33\pm0.21$ \nl
\enddata
\end{deluxetable}

\clearpage
\begin{deluxetable}{lccccc}
\tablecaption{IC 1613, SMC, and LMC Distance Moduli \label{tabSMCcomp}}
\tablehead{
\colhead{Method} &
\colhead{IC 1613} &
\colhead{SMC} &
\colhead{IC 1613 - SMC} &
\colhead{LMC} &
\colhead{IC 1613 - LMC}}
\startdata
RR Lyrae & $24.32 \pm 0.05$\tablenotemark{a} & $18.88 \pm 0.07$\tablenotemark{b} & $5.44 \pm 0.09$ & $18.49 \pm 0.07\tablenotemark{c}$ & $5.83 \pm 0.09$ \nl
Cepheid & $24.40 \pm 0.13$\tablenotemark{d} & $18.91 \pm 0.04$\tablenotemark{e} & $5.49 \pm 0.14$ & $18.50 \pm 0.00$\tablenotemark{f} & $5.90 \pm 0.13$ \nl
RGB Tip & $24.32 \pm 0.07$\tablenotemark{g} & $18.90 \pm 0.04$\tablenotemark{h} & $5.42 \pm 0.08$ & $18.57 \pm 0.09$\tablenotemark{i} & $5.82 \pm 0.11$ \nl
Red Clump & $24.30 \pm 0.09$ & $18.85 \pm 0.06$\tablenotemark{j} & $5.45 \pm 0.11$ & $18.46 \pm 0.11$\tablenotemark{k} & $5.84 \pm 0.14$ \nl
\enddata
\tablenotetext{a}{Adopting $M_V(RR) = 0.60$.  For this and all other distances in this table, calibration uncertainties are assumed to be zero because we are only interested in measuring relative distances.}
\tablenotetext{b}{From Walker \& Mack (1988) for NGC 121 ($\logz=-1.4$), adopting $M_V(RR) = 0.58 \pm 0.02$.}
\tablenotetext{c}{From Walker (1992) for clusters with mean $\logz=-1.9$, adopting $M_V(RR) = 0.45 \pm 0.06$.}
\tablenotetext{d}{From Madore \& Freedman (1991), adjusting to our extinction of $A_V = 0.08$.}
\tablenotetext{e}{Laney \& Stobie (1994), adopting an LMC distance modulus of 18.50}
\tablenotetext{f}{LMC distance modulus of 18.50 is assumed in calibration of IC 1613 and SMC distances}
\tablenotetext{g}{Adopting $M_I(TRGB)=-4.02$}
\tablenotetext{h}{From Cioni et al. (2000), adopting $M_I(TRGB)=-4.02$ and $A_I=0.07$ based on Schlegel et al. (1998)}
\tablenotetext{i}{From Sakai, Zaritsky, \& Kennicutt (2000), adopting $M_I(TRGB)=-4.03$ calculated from the Girardi et al. (2000) models}
\tablenotetext{j}{From Girardi \& Salaris (2000)}
\tablenotetext{k}{From Girardi \& Salaris (2000), adopting the mean of their two possible values}
\end{deluxetable}

\clearpage
\begin{deluxetable}{lccc}
\tablecaption{IC 1613 Distance Moduli \label{tabDist}}
\tablehead{
\colhead{Method} &
\colhead{This Work} &
\colhead{Primary source of error} &
\colhead{Best Value}}
\startdata
RR Lyrae & $24.32 \pm 0.16$ & $M_V$ vs. $\logz$ calibration & $24.32 \pm 0.16$ \nl
Cepheid & $24.45 \pm 0.15$ & Small number (2) & $24.34 \pm 0.20$ \nl
RGB Tip & $24.32 \pm 0.09$ & Measurement & $24.32 \pm 0.09$ \nl
Red Clump & $24.30 \pm 0.09$ & Age effects & $24.30 \pm 0.09$ \nl
Combined & & & $24.31 \pm 0.06$ \nl
\enddata
\end{deluxetable}

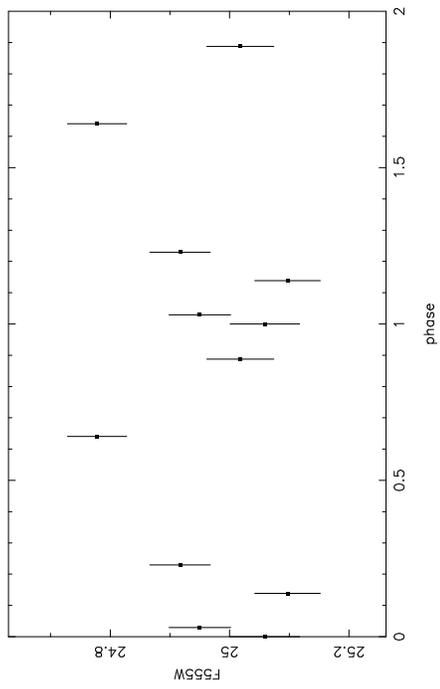
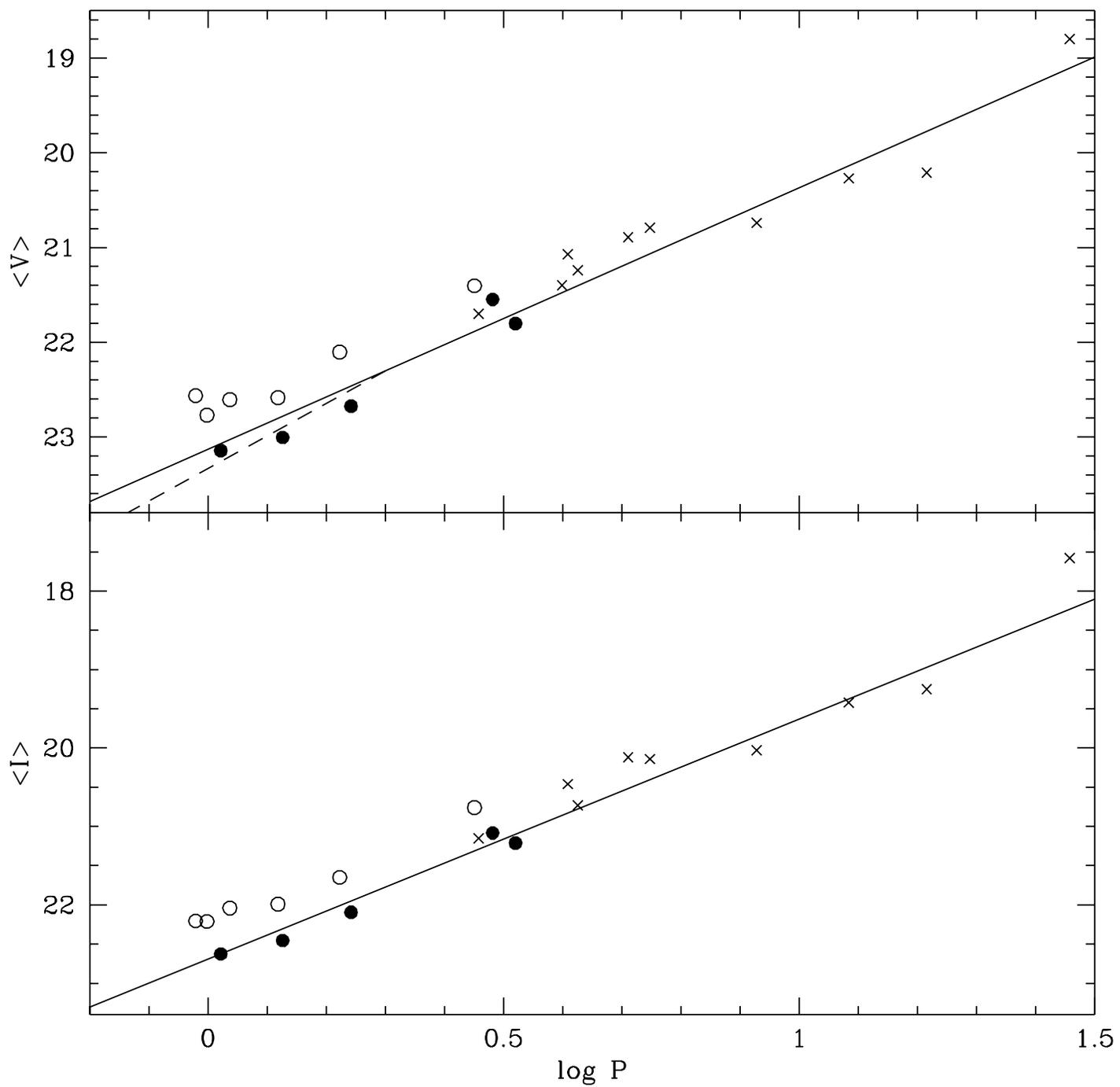
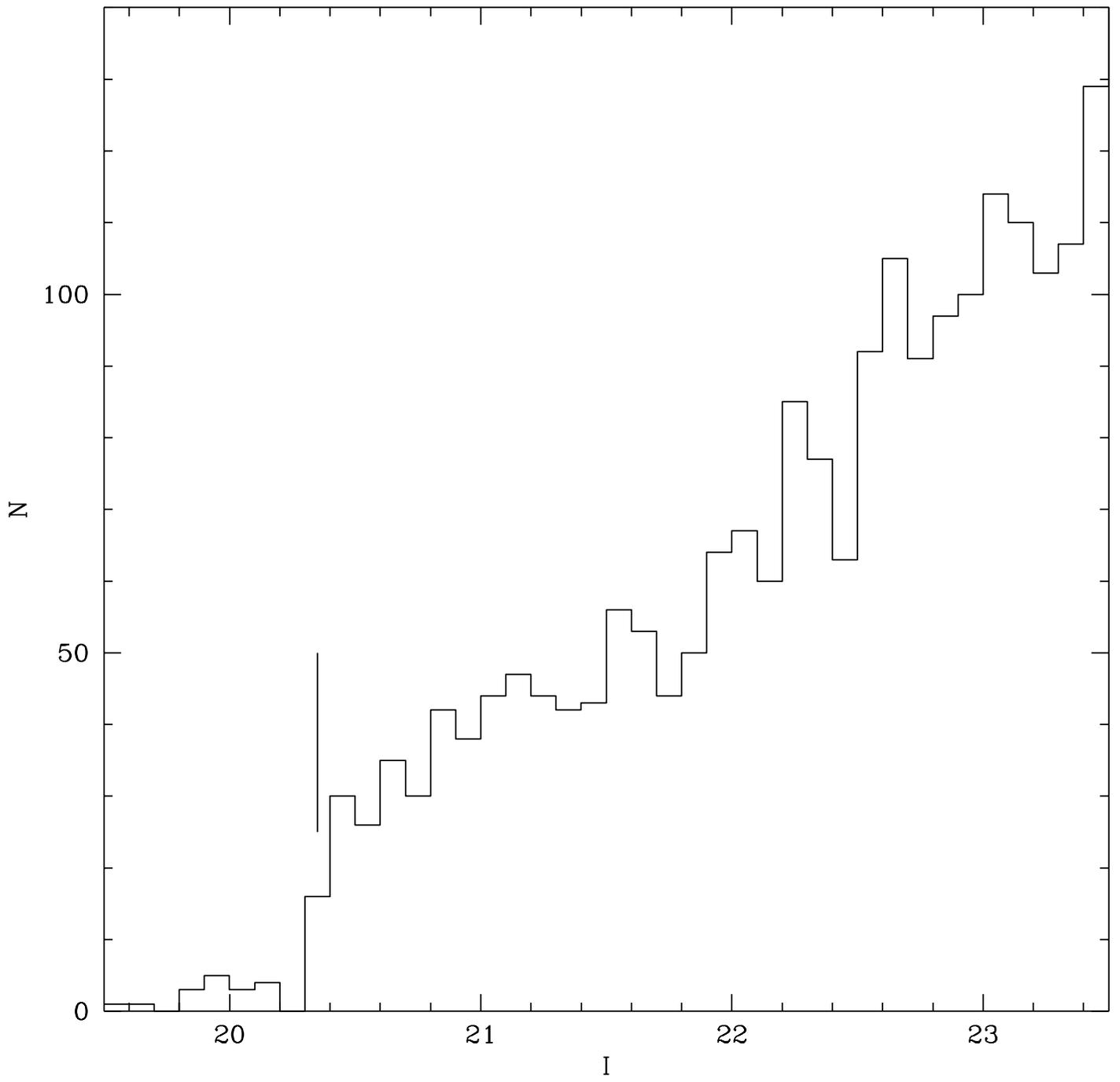
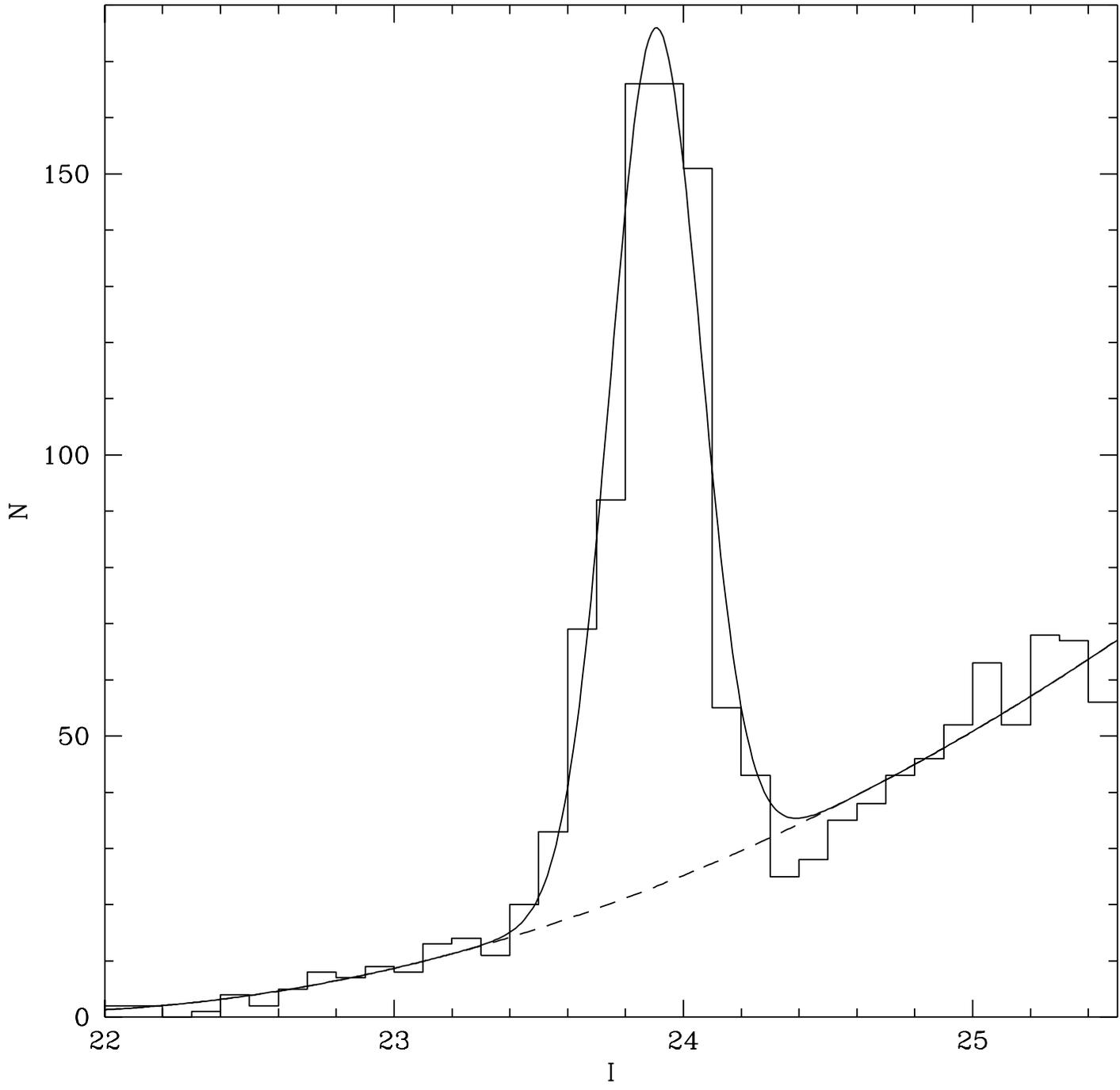
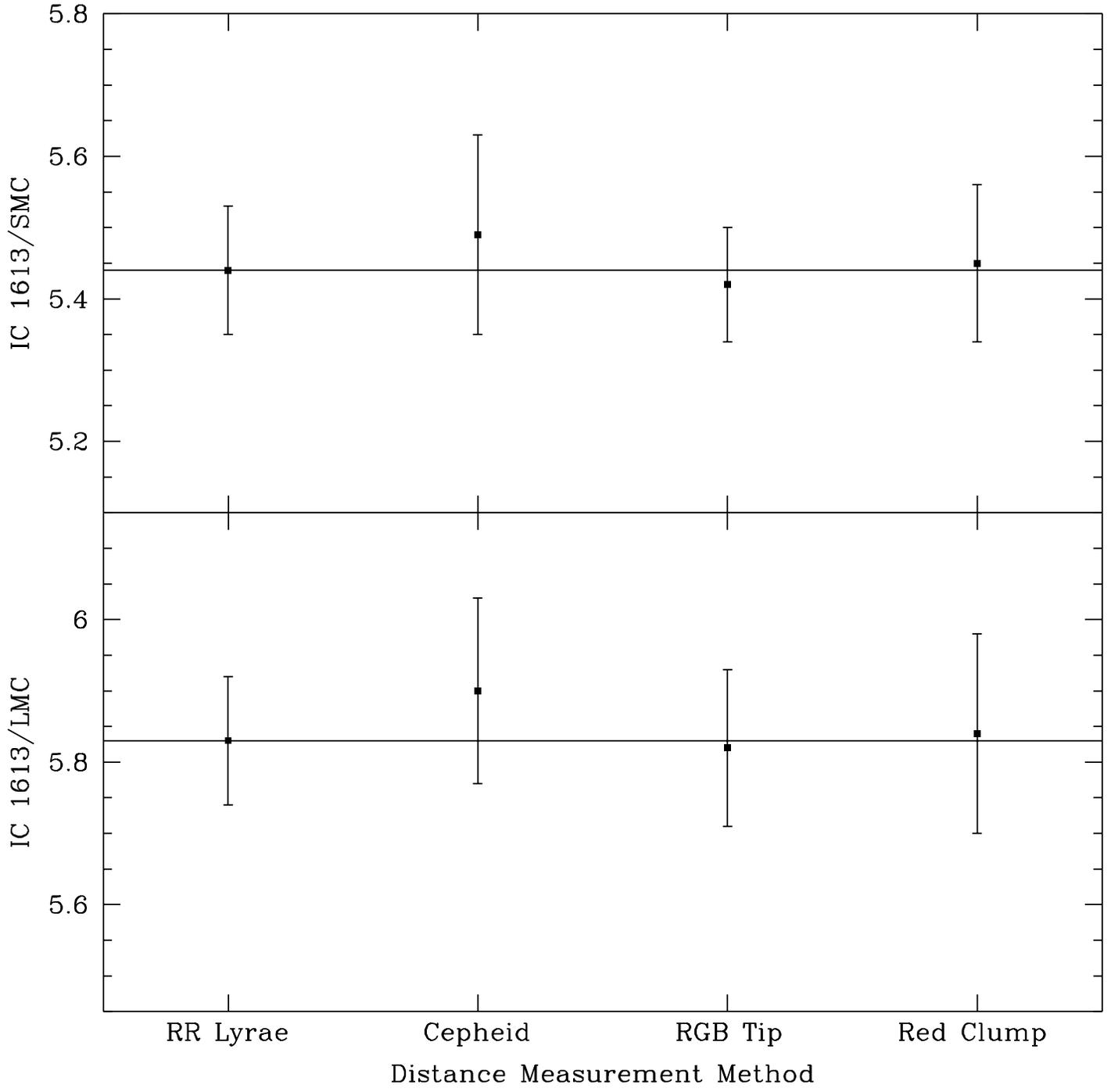
\end{document}